\documentclass[journal]{IEEEtran}

\usepackage{cite}
\usepackage{amsmath,amssymb,amsfonts}
\usepackage{graphicx}
\usepackage{textcomp}
\usepackage{xcolor}
\usepackage{booktabs}
\usepackage{multirow}
\usepackage{diagbox}
\usepackage{hyperref}
\hypersetup{
    colorlinks=true,
    linkcolor=blue,
    citecolor=blue,
    urlcolor=blue,
}

\newcommand{\etal}{\textit{et al.}}

\begin{document}

\title{JSR-GFNet: Jamming-to-Signal Ratio-Aware Dynamic Gating for
 Interference Classification in future Cognitive Global Navigation Satellite Systems}
\author{Zhihan Zeng, Hongyuan Shu, Kaihe Wang, Lu Chen, Amir Hussian, \IEEEmembership{Senior Member, IEEE},\\
Yanjun Huang, Junchu Zhao, Yue Xiu, \IEEEmembership{Member, IEEE}, Zhongpei Zhang
\thanks{Zhihan Zeng, Yue Xiu, Zhongpei Zhang are with National Key Laboratory of Wireless Communications, University of Electronic Science and Technology of China, Chengdu 611731, China (E-mail: 202511220608@std.uestc.edu.cn, xiuyue12345678@163.com, zhangzp@uestc.edu.cn). Kaihe Wang is with University of Electronic Science and Technology of China, Chengdu 611731, China (E-mail: khewang@yeah.net). Hongyuan Shu and Junchu Zhao are with Shanghai Aerospace Electronic Technology Institute, Shanghai 201109, China, and also with Shanghai Key Laboratory of Collaborative Computing in Spacial Heterogeneous Networks (CCSN), Shanghai 201109, China (E-mail: shuhongyuan2020@163.com, zhaojunchuhit@163.com). Lu Chen is with the Anhui Science and Technology University
Chuzhou 239000, China and Nanjing University Of Information Science $\&$ technology, Nanjing 210044, China (E-mail: chenlu@ahstu.edu.cn). Amir Hussian is with the University of Oxford, UK (E-mail: hussain.doctor@gmail.com). Yanjun Huang is with Shanghai Xiaoyuan Innovation center, China (E-mail: poer1985@163.com).
The corresponding author is Yanjun Huang.}}
\maketitle

\begin{abstract}
The transition toward cognitive global navigation satellite system (GNSS) receivers requires accurate interference classification to trigger adaptive mitigation strategies. However, conventional methods relying on Time-Frequency Analysis (TFA) and Convolutional Neural Networks (CNNs) face two fundamental limitations: severe performance degradation in low Jamming-to-Signal Ratio (JSR) regimes due to noise obscuration, and ``feature degeneracy'' caused by the loss of phase information in magnitude-only spectrograms. Consequently, spectrally similar signals---such as high-order Quadrature Amplitude Modulation versus Band-Limited Gaussian Noise---become indistinguishable. To overcome these challenges, this paper proposes the \textbf{JSR-Guided Fusion Network (JSR-GFNet)}. This multi-modal architecture combines phase-sensitive complex In-Phase/Quadrature (IQ) samples with Short-Time Fourier Transform (STFT) spectrograms. Central to this framework is a physics-inspired dynamic gating mechanism driven by statistical signal descriptors. Acting as a conditional controller, it autonomously estimates signal reliability to dynamically reweight the contributions of a Complex-Valued ResNet (IQ stream) and an EfficientNet backbone (STFT stream). To validate the model, we introduce the Comprehensive GNSS Interference (CGI-21) dataset, simulating 21 jamming categories including software-defined waveforms from aerial platforms. Extensive experiments demonstrate that JSR-GFNet achieves higher accuracy across the full 10--50 dB JSR spectrum. Notably, interpretability analysis confirms that the model learns a physically intuitive strategy: prioritizing spectral energy integration in noise-limited regimes while shifting focus to phase precision in high-SNR scenarios to resolve modulation ambiguities. This framework provides a robust solution for next-generation aerospace navigation security.
\end{abstract}

\begin{IEEEkeywords}
GNSS Interference Classification, Deep Learning, Multi-Modal Fusion, Complex-Valued Neural Networks, Jamming-to-Signal Ratio (JSR), Adaptive Gating.
\end{IEEEkeywords}

\section{Introduction}
\label{sec:intro}

Global Navigation Satellite Systems (GNSS) have evolved into a fundamental pillar of modern critical infrastructure, providing critical Positioning, Navigation, and Timing (PNT) services for sectors ranging from aviation and power grids to autonomous transportation. However, the inherent fragility of GNSS signals, which arrive at the Earth's surface with power levels significantly below the thermal noise floor, renders receivers exceptionally vulnerable to Radio Frequency Interference (RFI) \cite{borio2016swept, borowski2012detecting}. Whether stemming from unintentional sources, such as harmonics from communication systems, or malicious attacks including jammers mounted on Aerial Autonomous Vehicles (AAVs) \cite{sun2025multiparameter,11220909}, RFI can degrade positioning accuracy or cause a total loss of signal lock. Consequently, the transition from passive interference suppression to ``Cognitive Anti-Jamming'' has become a research priority.\cite{gao2025general} In this paradigm, the accurate classification of the interference type constitutes the prerequisite step, enabling the receiver to autonomously select the most effective mitigation strategy, such as activating notch filters for narrowband tones or pulse blanking for high-power bursts \cite{falletti2020design, hegarty2000suppression,11353414,11355857,11316665,11098592,11220909,11207524,11346858,11316633,11108293,10797657}.

The characterization of modern interference sources, however, presents distinct signal processing challenges that transcend traditional energy detection methods \cite{motella2014methods, wang2017time}. As threat emitters evolve from simple continuous wave generators to software-defined platforms capable of mimicking communication waveforms, the spectral and temporal signatures of jamming signals have become increasingly complex. This evolution introduces two fundamental hurdles for reliable classification. The first is the critical issue of \textit{Phase Ambiguity} and ``Feature Degeneracy.'' Many high-order Digital Modulation Interferences (DMI), such as 64-QAM, exhibit power spectral densities that are virtually indistinguishable from Band-Limited Gaussian Noise (BLGNI). While these signals possess fundamentally different constellation properties in the complex In-Phase/Quadrature (IQ) domain, standard Time-Frequency Analysis (TFA) methods—which typically rely on magnitude-only representations—permanently discard this vital phase information. The second hurdle is the extreme variability of the Jamming-to-Signal Ratio (JSR). In realistic operational scenarios, such as incipient jamming or long-distance standoff attacks, interference energy is often diffuse and submerged in the background noise floor, rendering signal features unstable and difficult to isolate \cite{jia2025low}.

To address these complexities, data-driven approaches, particularly Deep Learning (DL), have recently emerged as a dominant paradigm, significantly outperforming classical statistical tests \cite{ferre2019jammer, xu2020gps}. The majority of existing literature formulates interference classification as a Computer Vision problem \cite{swinney2021gnss}. By transforming one-dimensional time-domain signals into TFA images, typified by spectrograms generated via the Short-Time Fourier Transform (STFT), researchers have successfully deployed Convolutional Neural Networks (CNNs) to identify visual jamming patterns \cite{ebrahimimehr2022detection, mehr2025deep}. Despite their prevalence, these single-modality, vision-based approaches inherently inherit the limitations of the signal representations they utilize. Specifically, by relying solely on spectrograms, these models suffer from severe performance degradation in low-JSR regimes where visual contrast is obscured by noise \cite{vandermerwe2024optimal}. More critically, the neglect of raw phase information renders them incapable of resolving the ambiguity between spectrally similar signals (e.g., distinguishing high-order QAM from noise), a vulnerability that sophisticated jammers can exploit.

To overcome these physical and methodological limitations, it is hypothesized that a robust classification framework for cognitive receivers must be inherently multi-modal \cite{zhong2024tsfanet, xiao2025compound}. Such a framework should be capable of leveraging both the global spectral patterns of the time-frequency domain (for robustness against wideband variations) and the subtle phase characteristics of the time domain (for modulation precision). However, simple concatenation of features from both domains is suboptimal because the reliability of each modality varies dynamically with the jamming environment. Phase information in raw IQ data is easily corrupted by noise at low JSRs, whereas spectrograms provide better energy integration; conversely, at high JSRs, the IQ domain offers the precision required to resolve modulation ambiguity that spectrograms lack.

In this paper, we propose the JSR-Guided Fusion Network (JSR-GFNet), a deep learning architecture designed for robust GNSS interference classification. The core mechanism of JSR-GFNet is a decoupled dynamic gating module. By utilizing statistical descriptors, including Kurtosis and the Peak-to-Average Power Ratio (PAPR), as a proxy for the jamming environment, the network learns to autonomously arbitrate between the time-domain (IQ) and frequency-domain (STFT) streams. This allows the model to dynamically weight the contribution of each modality, ensuring optimal feature utilization across varying JSR levels.

The main contributions of this paper are summarized as follows:

1) \textit{Physics-Aware Multi-Modal Architecture:} We propose JSR-GFNet, which integrates a Complex-Valued ResNet for phase-sensitive IQ processing and an EfficientNet backbone for robust visual feature extraction. A unique JSR-guided gating mechanism is introduced to adaptively fuse these features, resolving the dilemma of modal reliability under varying noise conditions.

2) \textit{Resolution of Phase Ambiguity:} We explicitly address the classification challenge of spectrally similar signals. By preserving phase integrity through the IQ stream, the proposed model successfully distinguishes high-order digital modulations, including the discrimination between 16-QAM and 64-QAM, from noise-like interference, a task where traditional STFT-only models fail.

3) \textit{Comprehensive Dataset and Scenario:} We construct the Comprehensive GNSS Interference (CGI-21) dataset, simulating a realistic terrestrial threat scenario involving 21 types of interference, such as sophisticated digital modulations and frequency-agile chirps. To ensure statistical robustness, the dataset comprises over 880,000 samples, featuring 2,000 independent realizations for each category across 21 distinct JSR levels. This dataset provides a rigorous benchmark for evaluating classification performance under ``phase-masked'' attack conditions.

4) \textit{State-of-the-Art Performance and Interpretability:} Extensive experiments demonstrate that JSR-GFNet achieves higher accuracy across the full JSR range of 10 to 50 dB. Furthermore, we provide a physical interpretation of the learned gating weights, confirming that the network aligns with signal processing intuition by shifting focus from visual (STFT) features in noise-limited regimes to phase (IQ) features as the signal quality improves.

The remainder of this paper is organized as follows. Section \ref{sec:methodology} details the proposed JSR-GFNet architecture. Section \ref{sec:experimental_setup} describes the simulation scenario and dataset. Section \ref{sec:results} presents the performance analysis and comparison. Finally, Section \ref{sec:conclusion} concludes the paper.

\section{Related Work}
\label{sec:related_work}

The evolution of GNSS interference countermeasures has transitioned from classical statistical signal processing to data-driven learning paradigms. This section reviews the existing literature in two main categories: traditional detection and mitigation methods, and recent advancements based on Machine Learning (ML) and DL.

\subsection{Traditional Methods: Detection and Mitigation}
\label{subsec:traditional_methods}

Prior to the advent of widespread neural network applications, interference defense relied heavily on statistical theory and signal processing heuristics.

\subsubsection{Interference Mitigation Techniques}
Mitigation strategies are typically tailored to the time-frequency characteristics of the jamming signal. For pulsed interference such as Distance Measuring Equipment (DME) signals, \textit{Pulse Blanking} is the standard approach. Borio \cite{borio2016swept} and Hegarty \etal \cite{hegarty2000suppression} established the foundations of blanking nonlinearities in the time domain. More recently, Musumeci and Dovis \cite{musumeci2021comparison, musumeci2014use} extended this to transformed domains, demonstrating the efficacy of wavelet transforms for excision.

To address more complex waveforms like chirps, advanced transform-domain methods have been developed. Sun \etal \cite{sun2024fractional} proposed using the Fractional Fourier Transform (FrFT) to concentrate sweep energy, significantly outperforming traditional techniques. Similarly, Luo \etal \cite{luo2024zak} introduced a method based on the Zak Transform, which leverages signal sparsity in the delay-Doppler domain to effectively extract and mitigate chirp interference.

For continuous wave (CW) signals, \textit{Adaptive Notch Filtering (ANF)} remains dominant. Falletti \etal \cite{falletti2020design} and Borio \etal \cite{borio2008two, borio2020mitigation} provided comprehensive analyses of multi-pole notch filters. However, standard ANF struggles with varying power levels. To mitigate this, Song \etal \cite{song2025power} proposed an Adaptive Sparse Filtering (ASF) algorithm that senses the equivalent bandwidth of power-enhanced interference, dynamically adjusting filter parameters for optimal suppression. Additionally, matrix-theoretic approaches like the Karhunen–Loève decomposition \cite{sharifi2022efficient} and recent advancements in Non-negative Matrix Factorization (NMF) by da Silva \etal \cite{dasilva2023radio} demonstrate the potential of semi-blind separation techniques.

\subsubsection{Interference Detection Techniques}
Detection typically precedes mitigation. \textit{Pre-correlation} methods monitor statistical anomalies in the raw Intermediate Frequency (IF) samples. Techniques based on Automatic Gain Control (AGC) levels \cite{borowski2012detecting, thompson2011detection} and spectral analysis \cite{tani2008performance, wang2018gnss} effectively flag high-power jammers. Motella and Lo Presti \cite{motella2014methods} introduced goodness-of-fit tests for robust detection. To enhance sensitivity in low-SNR scenarios, Wang \etal \cite{wang2017time} proposed a statistical inference technique combining Pseudo-Wigner-Ville distributions with F-tests. Furthermore, recent works utilize advanced transforms such as the Radon-Wigner transform \cite{sun2023novel} to detect sweep interference. \textit{Post-correlation} methods, conversely, analyze receiver observables like $C/N_0$ \cite{falletti2011low, bhuiyan2014impact}, offering a complementary detection layer.

\subsection{Machine Learning and Deep Learning Approaches}
\label{subsec:ml_dl_methods}

The complexity of modern jamming waveforms has driven the shift towards data-driven classification, utilizing both Computer Vision and sequence learning paradigms. A dominant paradigm involves converting 1D signals into 2D spectrograms, treating classification as a Computer Vision task. Ferre \etal \cite{ferre2019jammer} and Mehr and Dovis \cite{ebrahimimehr2022detection} pioneered the use of CNNs on spectrograms, and more recently, Mehr and Dovis \cite{mehr2025deep} extended this work, validating CNN robustness across a wider range of jammer types. Specialized architectures have also emerged; for instance, Fu \etal \cite{fu2024research} adapted YOLOv8 for interference localization, while Zhang \etal \cite{zhang2023radar} applied object detection networks to radar jamming. Addressing the challenge of detection under low-power conditions, Jia \etal \cite{jia2025low} proposed a hierarchical CNN framework that fuses spectrograms with autocorrelation function maps and spectral flatness, achieving high accuracy even when interference is submerged in noise. To reduce computational overhead, researchers have utilized manual features with classical classifiers like SVMs \cite{xu2020gps, qin2022situational, vandermerwe2024optimal}, while in the deep learning domain, sequence models operating on raw IQ data have been proposed, such as LSTM networks \cite{fu2022navigation} and autocorrelation-based architectures like JRNet \cite{qu2020jrnet} and IRNet \cite{qu2022irnet}. Recent trends prioritize efficiency for edge deployment; Mehr \etal \cite{mehr2025faster} introduced a GRU-based approach that processes spectrogram slices sequentially to reduce latency, and Jia \etal \cite{jia2025lightweight} explored Kolmogorov-Arnold Networks (KAN) combined with multi-scale feature fusion for few-shot learning. As the threat landscape evolves beyond simple jamming, Xiao \etal \cite{xiao2025compound} tackled compound interference by integrating time-frequency features with power spectrum data, and Sun \etal \cite{sun2025multiparameter} addressed spoofing using a Time-Space Variational Autoencoder (TSVAE) for unsupervised detection. Deep learning is also entering the mitigation domain, with Song \etal \cite{song2024gnss} applying U-Net architectures for pixel-level interference segmentation. Although recent research posits that fusing time-domain and frequency-domain features yields superior robustness—as seen in frameworks like TSFANet \cite{zhong2024tsfanet} and attention-based networks \cite{cai2024attention}—a critical limitation persists: the fusion strategy is typically \textit{static}. Existing methods often concatenate features with fixed weights, ignoring that modal reliability fluctuates with JSR. Spectrograms excel in noise due to integration gain, while IQ phase data is crucial for resolving modulation ambiguity at high JSRs; a static fusion mechanism cannot adapt to these variations. To address these limitations, this paper proposes a multi-modal deep learning framework driven by a dynamic gating mechanism, designed to autonomously arbitrate between phase-sensitive IQ data and robust visual features based on the jamming environment, thereby ensuring optimal feature utilization across varying signal power regimes.

\section{Methodology}
\label{sec:methodology}

\begin{figure*}[htbp]
    \centering
    \includegraphics[width=0.75\linewidth]{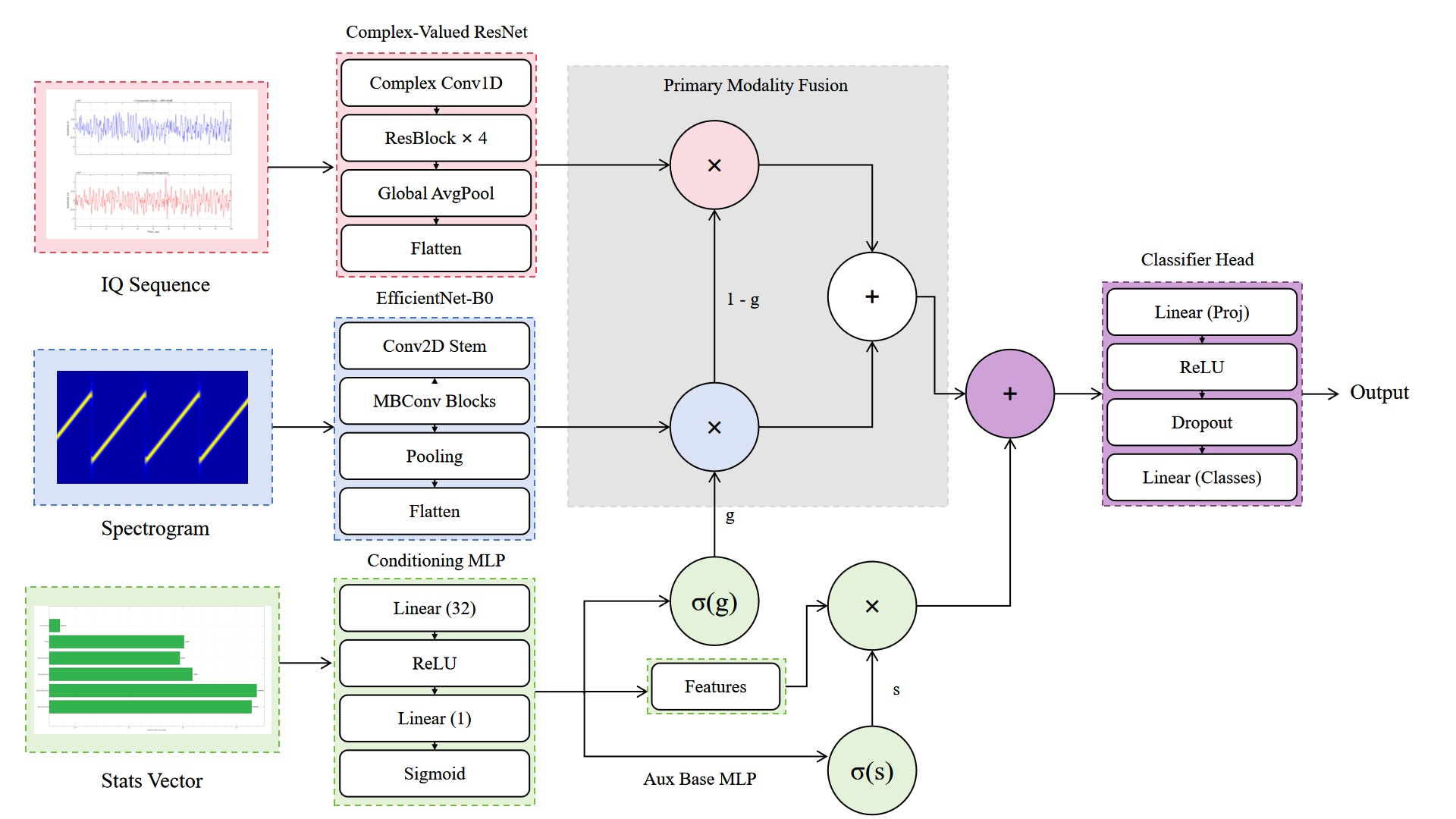}
    \caption{Schematic representation of the proposed JSR-GFNet architecture.}
    \label{fig:architecture}
\end{figure*}

GNSS are critical infrastructures for modern PNT services. However, due to the extremely low power of satellite signals upon reaching the Earth's surface, which are typically below the thermal noise floor, GNSS receivers are inherently vulnerable to RFI. Malicious jamming attacks, ranging from simple continuous wave tones to sophisticated frequency-agile waveforms, can easily saturate the receiver front-end or distort correlation peaks, leading to a complete loss of lock. While traditional energy detection techniques can effectively flag the presence of high-power interference, they often struggle to classify the specific type of jamming, which is a prerequisite for activating appropriate mitigation algorithms.

The complexity of modern jamming signals, particularly chirp-based and frequency-hopping modulations, necessitates a representation that captures time-varying spectral characteristics. TFA, such as the STFT, has emerged as a powerful tool for visualizing non-stationary signals as images (spectrograms), enabling the application of Computer Vision techniques. However, relying solely on spectrogram images has limitations, particularly in low JSR scenarios where the visual signature of the interference is masked by background noise. Conversely, time-domain IQ data preserves subtle phase information crucial for identifying digital modulations but lacks the global spectral context.

To address these challenges, this paper proposes a multi-modal deep learning framework named JSR-GFNet. Unlike existing methods that rely on a single input domain or static feature concatenation, JSR-GFNet introduces an adaptive gating mechanism. By utilizing statistical features as a proxy for the jamming environment, the network dynamically weighs the importance of time-domain (IQ) and time-frequency (STFT) representations. This allows the model to leverage robust visual features from spectrograms when the jamming signal is submerged in noise (low JSR), and seamlessly switch to precision phase-sensitive features from the IQ stream to resolve modulation ambiguities as the signal quality improves.

\subsection{Signal Model and Jamming Formulations}
The signal received at the GNSS receiver front-end, $r(t)$, is modeled as a superposition of the legitimate GNSS satellite signals $s(t)$, the jamming signal $j(t)$, and Additive White Gaussian Noise (AWGN) $n(t)$:
\begin{equation}
    r(t) = \sum_{i=1}^{M} s_i(t) + j(t) + n(t),
    \label{eq:signal_model}
\end{equation}
where $M$ denotes the number of visible satellites. To comprehensively evaluate the robustness of the classifier against diverse threat scenarios, we simulate 21 distinct jamming types based on their time-frequency characteristics. The mathematical formulations are categorized into specific groups as follows.

\subsubsection{Digital Modulation Interferences (DMI)}
This category emulates communication waveforms, which can be particularly deceptive to receivers. It encompasses BPSK, QPSK, and high-order QAM, specifically 8-, 16-, 32-, and 64-QAM. These signals are generated by mapping random data bits to complex symbols, followed by pulse shaping to constrain bandwidth. The unified model is:
\begin{equation}
    j_{\text{DMI}}(t) = \sqrt{P_J} \cdot \left[ \sum_{n=-\infty}^{\infty} d_n h(t - nT_{sym}) \right] e^{j(2\pi f_{\Delta} t + \phi_0)},
\end{equation}
where $P_J$ is the jamming power, $d_n$ represents the complex symbol drawn from the modulation constellation $\mathcal{A}$ (e.g., $\mathcal{A}_{\text{64QAM}}$), $T_{sym}$ is the symbol period, and $h(t)$ is the Root Raised Cosine (RRC) filter impulse response. Crucially, while distinct in the complex IQ domain, these signals often exhibit identical power spectral densities to Gaussian noise, leading to ambiguity in magnitude-only time-frequency representations. A formal proof of this fundamental non-injectivity of the spectrogram mapping is provided in Appendix A.

\subsubsection{Linear Frequency Modulation (LFM) Chirps}
Chirp jammers maximize their spectral efficiency by sweeping across the GNSS bandwidth. This group comprises \textit{LChirp} (with varying sweep rates) and \textit{SawChirp}. The instantaneous frequency changes linearly with time, creating a constant phase acceleration. The analytical signal is expressed as:
\begin{equation}
    j_{\text{LFM}}(t) = \sqrt{P_J} \exp\left[ j\left( 2\pi f_{\text{start}} t + \pi \frac{B}{T_{\text{swp}}} t^2 + \phi_0 \right) \right],
\end{equation}
where $B$ is the sweep bandwidth and $T_{\text{swp}}$ is the sweep period. For \textit{SawChirp}, the frequency resets instantaneously at the end of each period, creating a distinct vertical line in the spectrogram.

\subsubsection{Non-Linear Frequency Modulation (NLFM) Chirps}
Signals in this category, specifically \textit{SinChirp}, exhibit a continuous but non-linear frequency trajectory. The frequency follows a sinusoidal function, often used to evade simple notch filters:
\begin{equation}
    j_{\text{Sin}}(t) = \sqrt{P_J} \exp\left[ j\left( 2\pi f_c t + \beta \sin(2\pi f_m t) \right) \right],
\end{equation}
where $\beta$ is the modulation index controlling the bandwidth, and $f_m$ determines the rate of frequency fluctuation.

\subsubsection{Piecewise and Complex Chirps}
This category includes \textit{Triangular Chirp}, \textit{HookChirp}, and \textit{TickChirp}, where the frequency function $f(t)$ is defined piecewise. For instance, the \textit{TickChirp} combines a stationary dwell time with a fast frequency transition, creating a distinct ``tick'' shape in the spectrogram:
\begin{equation}
    f_{\text{Tick}}(t) =
    \begin{cases}
    f_k, & t \in [t_k, t_k + T_{\text{dwell}}] \\
    f_k + \alpha (t - t_{\text{trans}}), & \text{otherwise}
    \end{cases}.
\end{equation}

\subsubsection{Pulse Jamming (PJ)}
Modeled after DME interference, PJ consists of a train of high-power pulses. Specifically, DME signals are composed of Gaussian-shaped pulse pairs separated by a fixed interval. The mathematical formulation is given by:
\begin{equation}
    j_{\text{PJ}}(t) = \sqrt{P_J} \sum_{k} g_{\tau}(t - \tau_k) e^{j(2\pi f_c t + \phi_k)},
\end{equation}
where $g_{\tau}(t)$ represents the Gaussian pulse pair function with a pulse width typically around $\tau \approx 3.5 \mu s$ and a pulse interval of $\Delta t = 12 \mu s$, and $\tau_k$ denotes the random arrival times following a Poisson distribution.

\subsubsection{Frequency Hopping (FH)}
The carrier frequency jumps discretely across $N$ spectral channels. The signal remains stationary for a dwell time $T_{\text{dwell}}$ before switching, making it difficult to track with standard phase-locked loops:
\begin{equation}
    j_{\text{FH}}(t) = \sqrt{P_J} \sum_{k} \Pi_{T_{\text{dwell}}}(t - kT_{\text{dwell}}) e^{j(2\pi f_{k} t + \phi_k)},
\end{equation}
where $\Pi(\cdot)$ is the rectangular window function and $f_k$ is the random carrier frequency selected for the $k$-th hop.

\subsubsection{Continuous Wave Interference (CWI)}
CWI is characterized by energy concentration at a single frequency or a set of discrete frequencies. It is one of the most common forms of unintentional interference, typically arising from harmonics of other systems. The signal is modeled as:
\begin{equation}
    j_{\text{CWI}}(t) = \sqrt{P_J} \sum_{k=1}^{K} e^{j(2\pi f_k t + \phi_k)},
\end{equation}
where $K$ denotes the number of tones ($K=1$ for single-tone CWI).

\subsubsection{Band-Limited Gaussian Noise (BLGNI)}
This wideband interference is designed to raise the thermal noise floor of the receiver, masking the GNSS signal. It is generated by filtering complex white Gaussian noise $\eta(t)$ through a specific shaping filter, such as a Butterworth filter, denoted by its impulse response $h_{BP}(t)$:
\begin{equation}
    j_{\text{BLGNI}}(t) = \sqrt{P_J} \cdot [h_{BP}(t) * \eta(t)].
\end{equation}

\begin{table}[htbp]
    \caption{Summary of Simulated Interference Classes}
    \label{tab:interference_classes}
    \centering
    \begin{tabular}{@{}ll@{}}
    \toprule
    \textbf{Category} & \textbf{Interference Types} \\ \midrule
    \multirow{2}{*}{Digital Modulation} & BPSK, QPSK, 8-QAM, \\
      & 16-QAM, 32-QAM, 64-QAM \\ \midrule
    \multirow{2}{*}{Linear Chirps} & LChirp Wide (Slow/Med/Fast/Rapid), \\
      & LChirp Narrow, Sawtooth Chirp \\ \midrule
    \multirow{2}{*}{Non-Linear/Piecewise} & Sinusoidal, Hooked Sawtooth, \\
      & Triangular, Triangular Wave, Tick \\ \midrule
    \multirow{1}{*}{Pulsed} &  Pulse Jamming \\ \midrule
    \multirow{1}{*}{Hopping} & Frequency Hopping (FH) \\ \midrule
    \multirow{1}{*}{Continuous Wave} & Single-Tone CWI \\ \midrule
    \multirow{1}{*}{Noise-Like} & Band-Limited Gaussian Noise (BLGNI) \\ \bottomrule
    \end{tabular}
\end{table}

\subsection{Proposed Architecture: JSR-GFNet}
The overall architecture of the JSR-Guided Fusion Network is illustrated in Fig.~\ref{fig:architecture}. The model employs a multi-stream structure to process the IQ sequence, spectrogram, and statistical features in parallel, ensuring a holistic understanding of the signal characteristics.

\subsubsection{Complex-Valued ResNet (IQ Stream)}
Standard CNNs treat the real and imaginary parts of IQ data as independent channels, typically denoted as $N \times 2$, ignoring the inherent phase correlation between them. This approach often distorts the phase integrity critical for identifying modulations like QPSK and QAM. To mitigate this, we design a custom Complex-Valued Residual Network. The core component is the \textit{ComplexConv1d} layer. Let $W = W_r + jW_i$ be the complex filter and $h = h_r + jh_i$ be the complex input. The convolution is performed as:
\begin{equation}
    \mathcal{F}_{conv}(h) = (W_r * h_r - W_i * h_i) + j(W_r * h_i + W_i * h_r).
\end{equation}
This formulation preserves phase integrity throughout the feature extraction process. The encoder consists of four residual stages, transforming the raw $4000 \times 2$ sequence into a 1024-dimensional feature vector.

\subsubsection{EfficientNet-B0 Backbone (STFT Stream)}
For the visual modality, we utilize EfficientNet-B0 \cite{tan2019efficientnet} due to its superior trade-off between accuracy and computational efficiency. It features Inverted Residual Blocks (MBConv) equipped with Squeeze-and-Excitation (SE) optimization, which adaptively recalibrates channel-wise feature responses. The first convolutional layer (Stem) is adapted to accept single-channel grayscale inputs. This stream extracts high-level visual patterns, such as the slope of linear chirps or the sparsity of frequency hopping, outputting a 1280-dimensional feature vector.

\begin{figure*}[htbp]
    \centering
    \includegraphics[width=0.75\linewidth]{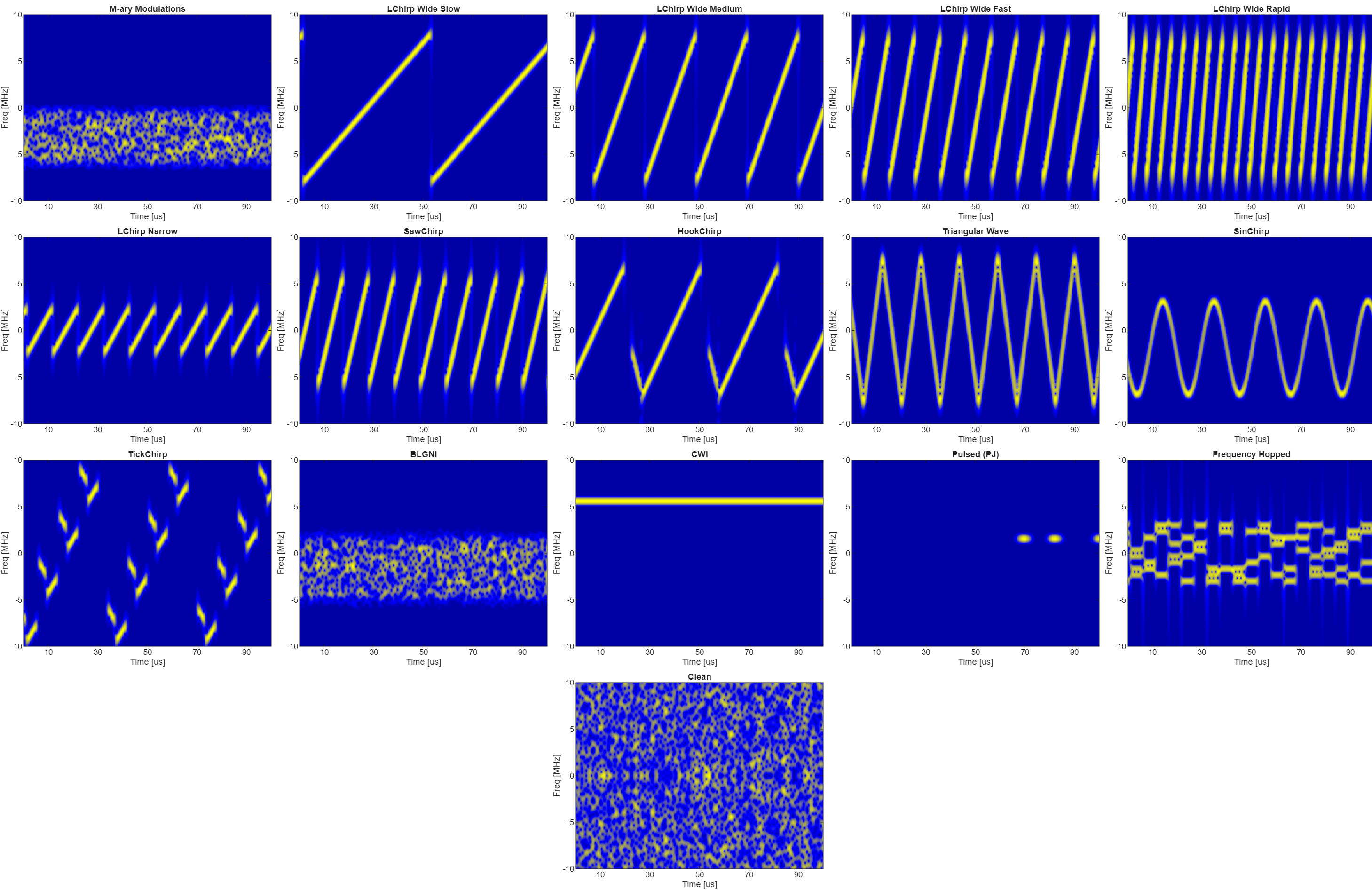}
    \caption{Spectrogram representations of the simulated jamming signals generated by STFT.}
    \label{fig:spectrogram_examples}
\end{figure*}

\subsubsection{Decoupled Dynamic Gating Mechanism}
Standard fusion methods often fail because the reliability of each modality varies significantly with the JSR. For instance, at low JSR, the spectrogram is dominated by noise, rendering visual features unreliable, while IQ statistical properties may still hold valid cues. This design is grounded in the theoretical insight that the discriminative information (e.g., Kullback-Leibler divergence) of IQ and STFT modalities follows opposite monotonic trends with respect to JSR, as formally derived in Appendix B.

We propose a decoupled gating strategy driven by the statistical branch. The network features a decoupled gating mechanism where statistical descriptors dynamically weight the contribution of time-domain (IQ) and time-frequency (Spectrogram) features based on the jamming intensity (JSR).

\textit{Primary Modality Selection ($g$):}
First, feature vectors from the IQ and STFT encoders are projected to a unified dimension (256-d) via Linear Projection layers to align the latent space. A learnable scalar gate $g \in [0, 1]$ is generated from the statistical vector:
\begin{equation}
    g = \sigma(\text{MLP}_{gate}(v_{stats})).
\end{equation}
The fused feature $Z_{fused}$ is obtained by weighted summation:
\begin{equation}
    Z_{fused} = g \cdot \text{Proj}_{stft}(z_{stft}) + (1-g) \cdot \text{Proj}_{iq}(z_{iq}).
\end{equation}
This mechanism enables ``soft-switching'': prioritizing phase-rich IQ features in low-SNR environments and texture-rich STFT features in high-SNR environments.

\textit{Auxiliary Feature Injection ($s$):}
To ensure critical statistical properties, including Kurtosis which is a strong indicator for pulse jamming, are not diluted during deep feature extraction, we employ an auxiliary gate $s$:
\begin{equation}
    Z_{final} = Z_{fused} + s \cdot \text{MLP}_{proc}(v_{stats}).
\end{equation}
The final feature vector $Z_{final}$ is passed to the classifier head for softmax probability generation.

\subsection{Data Preprocessing and Representations}
To leverage the complementary information from different domains, we preprocess the raw received signal $r(t)$ into three distinct modalities before feeding them into the neural network.

\subsubsection{Time-Frequency Representation (STFT)}
Given the non-stationary nature of jamming signals, the STFT is employed to capture time-varying spectral content. For a discrete signal $r[n]$, the STFT is defined as:
\begin{equation}
    S[m, k] = \sum_{n=0}^{N-1} r[n] w[n-m] e^{-j\frac{2\pi k n}{N}},
\end{equation}
where $w[n]$ is the Hamming window function ($L=256$, overlap=95\%). To enhance the visibility of low-power jamming features hidden in noise, we calculate the Power Spectral Density (PSD) $P = |S|^2$ and apply logarithmic mapping followed by adaptive min-max normalization:
\begin{equation}
    X_{\text{STFT}} = \frac{\log(P) - \min(\log(P))}{\max(\log(P)) - \min(\log(P))}.
\end{equation}
The resulting spectrogram is resized to $224 \times 224$ to match the visual encoder's input requirements. Representative spectrogram examples for different jamming categories are illustrated in Fig.~\ref{fig:spectrogram_examples}.

\begin{figure*}[htbp]
    \centering
    \includegraphics[width=0.95\linewidth]{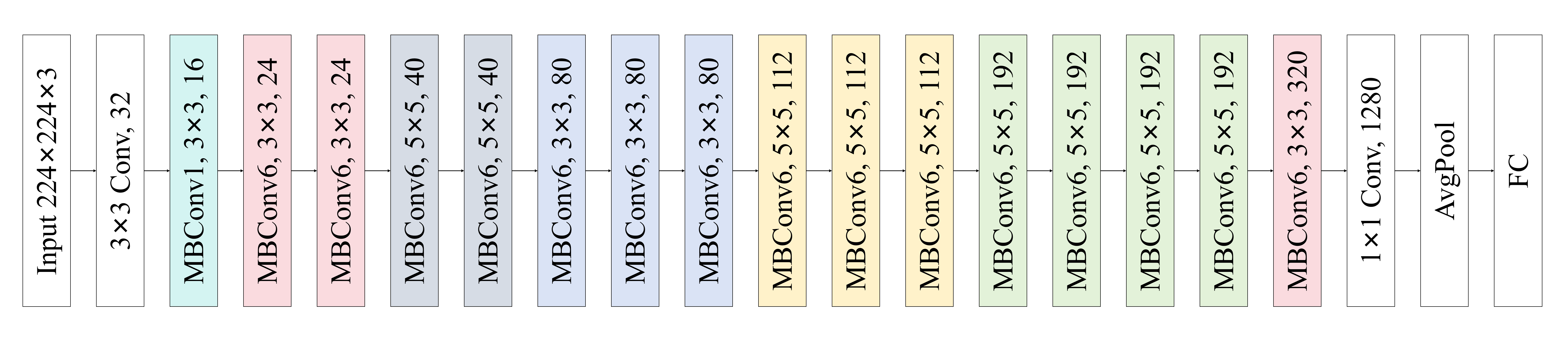}
    \caption{Detailed architecture of the EfficientNet-B0 backbone, featuring the stack of MBConv blocks and the final classification head.}
    \label{fig:efficientnet_arch}
\end{figure*}

As shown in Fig.~\ref{fig:spectrogram_examples}, different classes exhibit distinct visual signatures: linear chirps show sloped lines, frequency hopping appears as discrete blocks, and digital modulations appear as noise-like bands.

\subsubsection{Statistical Feature Extraction}
While deep learning extracts implicit features, explicit statistical descriptors provide robust global context. We compute a feature vector $v_{stats} \in \mathbb{R}^6$ comprising spectral features (Centroid, Bandwidth, Kurtosis, Flatness) and time-domain features (PAPR, Envelope Standard Deviation). These metrics effectively distinguish modulation types, including constant versus non-constant envelope signals, and identify impulsive noise behaviors.

\subsection{Convolutional Neural Networks}
To rigorously evaluate the performance of our proposed multi-modal framework, we benchmark it against several state-of-the-art deep learning architectures. Deep learning techniques, particularly CNNs, have revolutionized pattern recognition tasks by emulating the neural connectivity of the biological brain. A typical CNN architecture is composed of a sequence of layers that transform the input volume into an output class score. Before detailing the specific baseline architectures used in this study, we briefly describe the fundamental components that constitute these networks:

\begin{enumerate}
    \item \textit{Convolutional Layer:} This is the fundamental engine for feature extraction. It utilizes learnable kernels to slide over the input data, performing element-wise multiplication and summation to generate feature maps. This operation allows the network to detect local patterns such as edges in early layers and complex shapes in deeper layers.
    \item \textit{Activation Function:} Applied after convolution, activation functions introduce non-linearity to the network, enabling the learning of complex decision boundaries. The Rectified Linear Unit (ReLU), defined as $f(x) = \max(0, x)$, is commonly used to accelerate convergence by suppressing negative values.
    \item \textit{Pooling Layer:} This layer performs down-sampling operations, such as Max Pooling, to reduce the spatial dimensionality of feature maps. This process decreases the number of parameters and computational complexity while making the representation more invariant to small translations and distortions.
    \item \textit{Fully Connected (FC) Layer:} Located at the end of the network, FC layers flatten the high-level feature maps into a vector. They map the extracted features to the final classification space, typically culminating in a Softmax function that outputs probability distributions over the jamming classes.
\end{enumerate}

In the following subsections, we describe the specific CNN architectures and the Vision Transformer selected as baselines for our comparative analysis.

\subsubsection{EfficientNet-B0}
EfficientNet \cite{tan2019efficientnet} represents a modern family of models optimized for both accuracy and efficiency through a compound scaling method that uniformly scales network width, depth, and resolution. We employ EfficientNet-B0, the baseline model of this family.

The detailed architecture is depicted in Fig.~\ref{fig:efficientnet_arch}. It begins with a standard $3 \times 3$ convolution stem, followed by a sequence of 7 Mobile Inverted Bottleneck (MBConv) blocks. Each block utilizes depthwise separable convolutions to reduce parameter count and integrates Squeeze-and-Excitation (SE) optimization to adaptively recalibrate channel-wise feature responses. This design allows the network to effectively highlight subtle spectral patterns in jamming signals while maintaining computational lightness. The network terminates with a $1 \times 1$ convolution layer, a global average pooling layer, and a fully connected layer for final classification.

\subsubsection{AlexNet}
AlexNet \cite{krizhevsky2012imagenet} is a pioneering deep convolutional neural network that demonstrated the power of deep learning for large-scale image classification. It consists of five convolutional layers followed by three fully connected layers. Key innovations introduced by AlexNet include the use of ReLU activation functions to accelerate training and Dropout layers to prevent overfitting. While historically significant, its large parameter size and relatively simple architecture often serve as a baseline for measuring the efficiency gains of modern networks.

\subsubsection{ResNet-18}
As neural networks become deeper, they often suffer from the vanishing gradient problem, leading to performance degradation. The Residual Network (ResNet) architecture \cite{he2016deep} addresses this by introducing "skip connections" that allow gradients to flow more easily during backpropagation. We utilize ResNet-18, an 18-layer variant that balances depth and computational cost. It is composed of residual blocks containing $3 \times 3$ convolutions and batch normalization layers. This architecture effectively captures hierarchical spatial features from spectrograms without the optimization difficulties associated with plain deep networks.

\subsubsection{Vision Transformer (ViT-B-16)}
To explore non-convolutional approaches, we include the Vision Transformer (ViT) \cite{dosovitskiy2020image}. Unlike CNNs, ViT treats an image as a sequence of patches and processes them using standard Transformer encoders originally designed for natural language processing. The ViT-B-16 variant splits the $224 \times 224$ spectrogram into $16 \times 16$ patches. By leveraging the self-attention mechanism, ViT can capture long-range global dependencies across the entire time-frequency plane, offering a contrasting perspective to the local receptive fields of CNNs.

\section{Experimental Setup}
\label{sec:experimental_setup}

To comprehensively evaluate the robustness and generalization capability of the proposed JSR-GFNet, we established a high-fidelity simulation framework. This section details the operational scenario, the construction of the specific dataset employed, the implementation details of the training phase, and the metrics used for performance assessment.

\subsection{Operational Scenario and Signal Model}

\begin{figure}[htbp]
    \centering
    \includegraphics[width=0.95\linewidth]{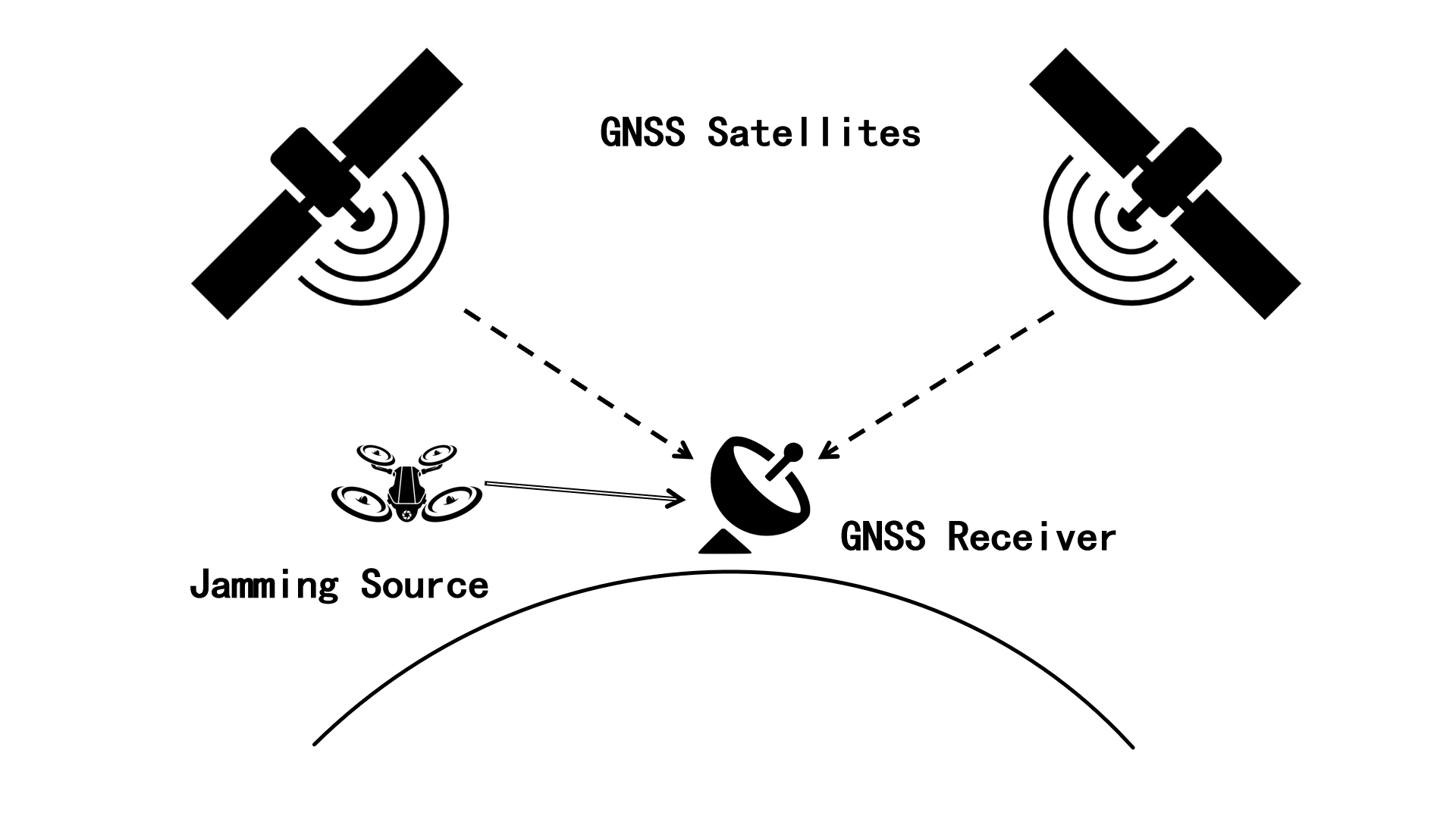}
    \caption{Illustration of the terrestrial interference monitoring scenario involving an Aerial Autonomous Vehicle (AAV) threat.}
    \label{fig:scenario}
\end{figure}

We consider a terrestrial interference monitoring scenario, as illustrated in Fig.~\ref{fig:scenario}. The system comprises a standard GNSS receiver located on the ground, whose legitimate communication link with satellite constellations is threatened by an intentional jamming source. Unlike traditional stationary jammers that emit simple continuous waves, we focus on a highly dynamic threat model involving an AAV.

The AAV is assumed to be equipped with SDR capable of generating sophisticated, frequency-agile interference. A particular focus of this study is on DMI, such as QAM. These signals are spectrally efficient and statistically similar to legitimate communication waveforms, making them significantly harder to distinguish from background noise compared to standard chirp or tonal jamming.

The received signal at the IF stage is modeled as a superposition of the legitimate GNSS signal, the interference signal, and AWGN. The simulation parameters are aligned with the GPS L1 C/A standard. The signal is sampled at a frequency of 40 MHz. Each data sample represents a snapshot of 100 microseconds, resulting in a sequence length of 4000 complex points. The GNSS signal power is set to -157 dBW, and the thermal noise floor is set to -205 dBW/Hz, creating a realistic, noise-dominated environment before jamming is introduced.

\subsection{The CGI-21 Dataset Construction}

To overcome the limitations of existing public datasets, which often lack diversity in digital modulation types, we constructed the CGI-21 dataset using a MATLAB-based simulation engine. The dataset encompasses 21 distinct interference categories, providing a broader coverage than typical 5-class or 12-class datasets found in the literature.

\subsubsection{Signal Parameters}
The specific parameters for all 21 interference types are detailed in Table~\ref{tab:sim_params}. The dataset includes six types of Digital Modulations (BPSK, QPSK, and 8-QAM up to 64-QAM), various LFM chirps with different sweep rates, non-linear chirps including Sawtooth, Triangular, and Tick waveforms, and pulsed interference.

\begin{table}[htbp]
    \caption{Parameters of Simulated Interference Signals in CGI-21}
    \label{tab:sim_params}
    \centering
    \renewcommand{\arraystretch}{1.3} 
    \resizebox{\columnwidth}{!}{
    \begin{tabular}{@{}llcc@{}}
    \toprule
    \multicolumn{2}{l}{\textbf{Interference Type}} & \textbf{Bandwidth} & \textbf{Key Characteristic} \\
    \multicolumn{2}{l}{} & \textbf{(MHz)} & \textbf{(per 100 $\mu$s Snapshot)} \\
    \midrule
     & BPSK, QPSK & $\approx 6.75$ & Symbol Rate = 5 Msps \\
     & 8-QAM, 16-QAM & $\approx 6.75$ & Roll-off Factor $\beta = 0.35$ \\
     & 32-QAM, 64-QAM & $\approx 6.75$ & Random Symbols \\
    \midrule
     & LChirp (Wide Slow) & 16 & Sweep Rate = 2 \\
     & LChirp (Wide Medium) & 16 & Sweep Rate = 5 \\
     & LChirp (Wide Fast) & 16 & Sweep Rate = 10 \\
     & LChirp (Wide Rapid) & 16 & Sweep Rate = 15 \\
     & LChirp (Narrow) & 5 & Sweep Rate = 10 \\
    \midrule
     & SawChirp & 12 & Sawtooth (Rate=11) \\
     & Triangular Wave & 16 & Triangular (Rate=10) \\
     & SinChirp & 10 & Sinusoidal Mod. (Rate=5) \\
     & HookChirp & $\approx 8$ & Piecewise Linear (Hook) \\
     & TickChirp & Variable & Quadratic + Hopping \\
    \midrule
     & BLGNI & 3 & 8th-order Butterworth Filter \\
     & Continuous Wave (CWI) & N/A & Single Tone (Random $f_c$) \\
    \midrule
     & Pulsed Jamming (PJ) & N/A & Gaussian Pulses (Rand arrival) \\
     & Frequency Hopped (FH) & 6 & Dwell Time = 5 $\mu$s \\
    \bottomrule
    \end{tabular}%
    }
    \vspace{1mm}
\end{table}

\subsubsection{Randomization Strategy}
To ensure the neural network learns intrinsic signal features rather than overfitting to synthetic artifacts, a rigorous randomization strategy was implemented. For every generated sample, a unique random seed is computed based on the class index, JSR, and sample index. This mechanism ensures that the phase offsets, symbol sequences for QAM, pulse arrival times, and the background AWGN realization are statistically independent across the entire dataset. The jamming intensity is varied to create a JSR ranging from 10 dB to 50 dB in 2 dB increments, providing a granular view of signal strength variations.

To leverage the multi-modal architecture of JSR-GFNet, the raw data is processed into three parallel streams. First, the IQ sequence is normalized to zero mean and unit variance to stabilize the gradients in the complex-valued residual blocks. Second, for the visual branch, we apply the STFT using a Hamming window of length 256 with 95\% overlap. The resulting power spectral density is log-transformed and resized to $224 \times 224$ pixels to fit the EfficientNet backbone. Finally, a vector of six statistical descriptors is extracted to guide the gating mechanism.

\subsection{Training Phase and Implementation Details}

Given the computational complexity of processing multi-modal data, the training process was executed on a high-performance computing platform. The detailed configuration regarding the computing environment, dataset partitioning, and optimization protocols is structured as follows.

\subsubsection{Computing Environment}
The deep learning models were implemented using the PyTorch library (version 2.8.0) in the Python 3.11 environment. All training and inference tasks were accelerated using an NVIDIA 5060 Ti GPU to handle the parallel computation requirements of the complex-valued convolutions.

\subsubsection{Dataset Partitioning Strategy}
To ensure a robust evaluation, we adopted a Mixed-Training and Stratified-Testing strategy. The training set is constructed by pooling 1,000 samples per class for each of the 21 JSR levels. This results in a massive, diverse training pool of 441,000 samples. This approach prevents the model from biasing towards specific power levels during the learning phase. For validation and testing, we reserved 20\% of the training data for monitoring overfitting, while the testing set is strictly separated and organized into specific JSR buckets. This test set comprises 220,500 samples in total, allowing for granular performance analysis at each signal-to-noise level.

\subsubsection{Optimization Protocol}
The network parameters are optimized using the Adam algorithm with a batch size of 64. The objective function is the Categorical Cross-Entropy Loss. To facilitate the convergence of the complex-valued residual blocks, we employ a composite learning rate scheduler spanning 30 epochs.

The learning rate schedule consists of two distinct phases. Initially, a warmup phase is applied for the first 3 epochs, where the learning rate increases linearly from 1\% of the base rate to the initial value of $5 \times 10^{-4}$. This is critical for stabilizing the early training dynamics of the multi-modal fusion layers. Subsequently, for the remaining 27 epochs, a cosine annealing schedule is adopted, allowing the learning rate to decay following a cosine curve. This strategy helps the model settle into a wider and more robust local minimum. Regularization is further applied via a weight decay of $1 \times 10^{-5}$ and a dropout rate of 0.5 in the fully connected layers.

\subsection{Evaluation Metrics}

To rigorously quantify the classification performance and analyze the model's behavior under varying interference conditions, we employ three complementary metrics: the Confusion Matrix, Overall Accuracy, and F1-Score.

\subsubsection{Confusion Matrix}
The Confusion Matrix serves as the foundational tool for our error analysis. It provides a tabular visualization of the discrepancies between the predicted classes and the ground truth labels. This metric is particularly crucial for this study as it reveals the specific "confusion patterns" among spectrally similar signals. For instance, distinguishing between different orders of Quadrature Amplitude Modulation, such as 16-QAM and 64-QAM, is inherently difficult due to their similar wideband noise-like signatures in the time-frequency domain.

Table~\ref{tab:confusion_matrix_structure} illustrates the structure of a binary confusion matrix using a diagonal split layout, which forms the conceptual basis for our multi-class evaluation.

\begin{table}[htbp]
    \caption{Structure of the Confusion Matrix}
    \label{tab:confusion_matrix_structure}
    \centering
    \renewcommand{\arraystretch}{1.5} 
    \begin{tabular}{|c|c|c|}
    \hline
    \diagbox[width=8em]{\textbf{Reality}}{\textbf{Prediction}} & \textbf{Anomaly} & \textbf{Normal} \\
    \hline
    \textbf{Anomaly} & True Positive (TP) & False Negative (FN) \\
    \hline
    \textbf{Normal} & False Positive (FP) & True Negative (TN) \\
    \hline
    \end{tabular}
\end{table}

By analyzing the off-diagonal elements (FP and FN) of the matrix, we can qualitatively assess whether the proposed multi-modal fusion successfully resolves ambiguities compared to single-modality baselines.

\subsubsection{Overall Accuracy}
Overall Accuracy (OA) provides a global measure of the model's effectiveness. It is defined as the ratio of correctly classified samples to the total number of samples in the dataset. Mathematically, it is expressed as:
\begin{equation}
    \text{Accuracy} = \frac{\sum_{i=1}^{C} TP_i}{N} \times 100\%,
\end{equation}
where $TP_i$ (True Positives) denotes the number of correctly identified samples for class $i$, and $N$ is the total number of test samples. While accuracy is a useful high-level indicator, it may not fully capture the performance nuances when specific classes are more challenging to detect than others, necessitating the use of more granular metrics like the F1-Score.

\subsubsection{F1-Score}
To provide a balanced assessment that considers both the model's ability to identify relevant signals (Recall) and its reliability in making positive predictions (Precision), we utilize the F1-Score. This metric is especially important when analyzing the model's performance on the DMI group at low JSR, where false negatives are common.

The F1-Score is calculated as the harmonic mean of Precision and Recall. First, for each class $i$, Precision ($P_i$) and Recall ($R_i$) are computed as:
\begin{equation}
    P_i = \frac{TP_i}{TP_i + FP_i}, \quad R_i = \frac{TP_i}{TP_i + FN_i},
\end{equation}
where $FP_i$ (False Positives) represents samples incorrectly assigned to class $i$, and $FN_i$ (False Negatives) represents samples of class $i$ that were missed. The F1-Score for class $i$ is then given by:
\begin{equation}
    \text{F1}_i = 2 \times \frac{P_i \times R_i}{P_i + R_i}.
\end{equation}
In our results section, we report the macro-averaged F1-Score across specific interference groups to highlight the model's robustness against complex, communication-like jamming signals.

\section{Results and Analysis}
\label{sec:results}

In this section, we present a comprehensive performance evaluation of the proposed JSR-GFNet. The evaluation metrics include OA, CM, and the F1-Score. We analyze the model's behavior from three perspectives: comparative performance against single-modality baselines, the impact of different visual backbones, and the interpretability of the gating mechanism.

\subsection{Overall Classification Performance}

To validate the effectiveness of the multi-modal fusion strategy, we first compared JSR-GFNet against representative single-modality baselines: an \textit{IQ-Only} model (Complex-ResNet) and an \textit{STFT-Only} model (EfficientNet-B0).

\begin{figure}[htbp]
    \centering
    \includegraphics[width=0.95\linewidth]{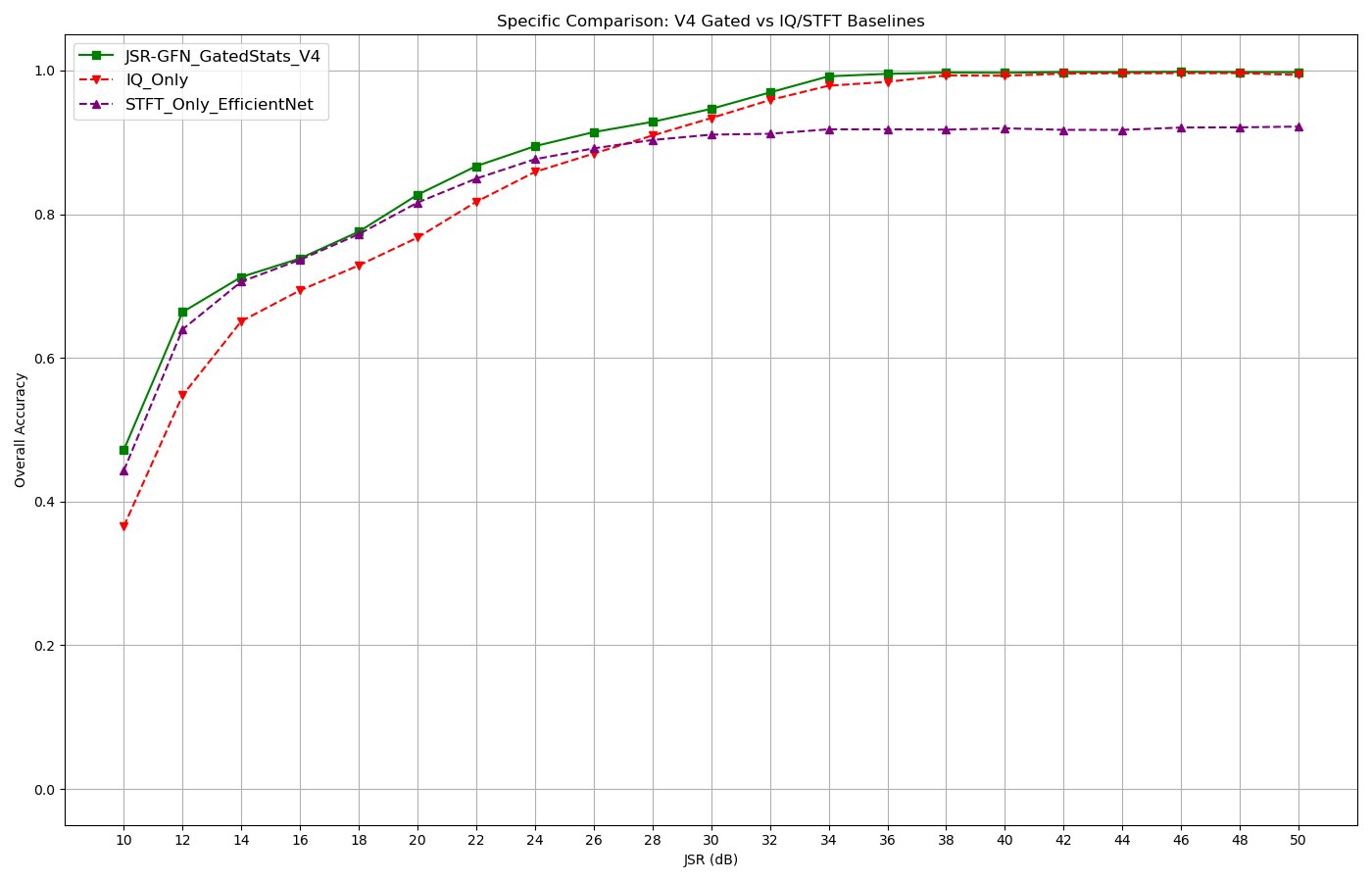}
    \caption{Overall classification accuracy comparison of JSR-GFNet against single-modality baselines.}
    \label{fig:overall_comparison}
\end{figure}

Fig.~\ref{fig:overall_comparison} illustrates the classification accuracy across the JSR spectrum. The results reveal a distinct ``crossing point'' in the baseline performances. In the Noise-Limited Regime (JSR $<$ 24 dB), the STFT-Only model (purple dashed line) significantly outperforms the IQ-Only model. For instance, at 10 dB, STFT-Only achieves much higher accuracy than the IQ-Only approach. This confirms that in high-noise environments, the energy integration property of spectrograms offers more robust features than raw IQ samples. Conversely, in the High-SNR Regime (JSR $>$ 30 dB), the IQ-Only model excels as the signal power increases, approaching 99.9\% accuracy. In contrast, the STFT-Only model saturates at a lower ceiling (approximately 92\%), likely due to the loss of fine-grained phase information inherent in the magnitude-only spectrogram generation.

Crucially, the proposed JSR-GFNet (green solid line) consistently tracks the upper envelope of both baselines, effectively avoiding the limitations of either single domain.

\subsection{Ablation Study: Visual Backbone Selection}

To justify the selection of EfficientNet-B0 as the visual encoder for the STFT branch, we conducted an ablation study comparing it against other standard architectures, including ResNet-18, ViT-B-16 (Vision Transformer), and AlexNet. Table~\ref{tab:backbone_comparison} summarizes the classification accuracy at various JSR levels derived from our experiments.

The results demonstrate that the JSR-GFNet (EfficientNet) configuration achieves the best overall performance, particularly in the challenging low-JSR regime (10-18 dB). While heavier models like ViT-B-16 show competitive performance at high JSRs, they suffer from convergence difficulties and lower accuracy when the spectrogram features are corrupted by heavy noise. EfficientNet-B0 provides the optimal balance between feature extraction capability and robustness for jamming recognition tasks.

\begin{table*}[htbp]
    \caption{Classification Accuracy (\%) Comparison of Different Visual Backbones (10-50 dB)}
    \label{tab:backbone_comparison}
    \centering
    \resizebox{\textwidth}{!}{
    \begin{tabular}{@{}lccccccccccc@{}}
    \toprule
    \multirow{2}{*}{\textbf{Backbone Architecture}} & \multicolumn{11}{c}{\textbf{Jamming-to-Signal Ratio (JSR)}} \\ \cmidrule(l){2-12}
      & \textbf{10 dB} & \textbf{14 dB} & \textbf{18 dB} & \textbf{22 dB} & \textbf{26 dB} & \textbf{30 dB} & \textbf{34 dB} & \textbf{38 dB} & \textbf{42 dB} & \textbf{46 dB} & \textbf{50 dB} \\ \midrule
    AlexNet & 36.9 & 70.2 & 76.1 & 82.8 & 90.1 & 94.2 & 99.1 & 99.8 & 99.9 & 100.0 & 100.0 \\
    ResNet-18 & 41.8 & 70.5 & 77.2 & 86.1 & 90.5 & 94.8 & 99.3 & 99.9 & 100.0 & 100.0 & 100.0 \\
    ViT-B-16 & 44.1 & 71.0 & 76.3 & 82.9 & 90.1 & 94.9 & 99.3 & 99.9 & 100.0 & 100.0 & 100.0 \\
    \textbf{EfficientNet-B0 (Ours)} & \textbf{47.2} & \textbf{71.5} & \textbf{78.1} & \textbf{86.5} & \textbf{91.3} & \textbf{95.1} & \textbf{99.5} & \textbf{100.0} & \textbf{100.0} & \textbf{100.0} & \textbf{100.0} \\ \bottomrule
    \end{tabular}
    }
\end{table*}

\subsection{Verification of the Gating Mechanism}

The core contribution of JSR-GFNet is its ability to dynamically weight modalities based on the jamming intensity. To verify this, we extracted the internal gate values ($g$) for all test samples, where $g$ represents the weight assigned to the STFT branch.

\begin{figure}[htbp]
    \centering
    \includegraphics[width=0.95\linewidth]{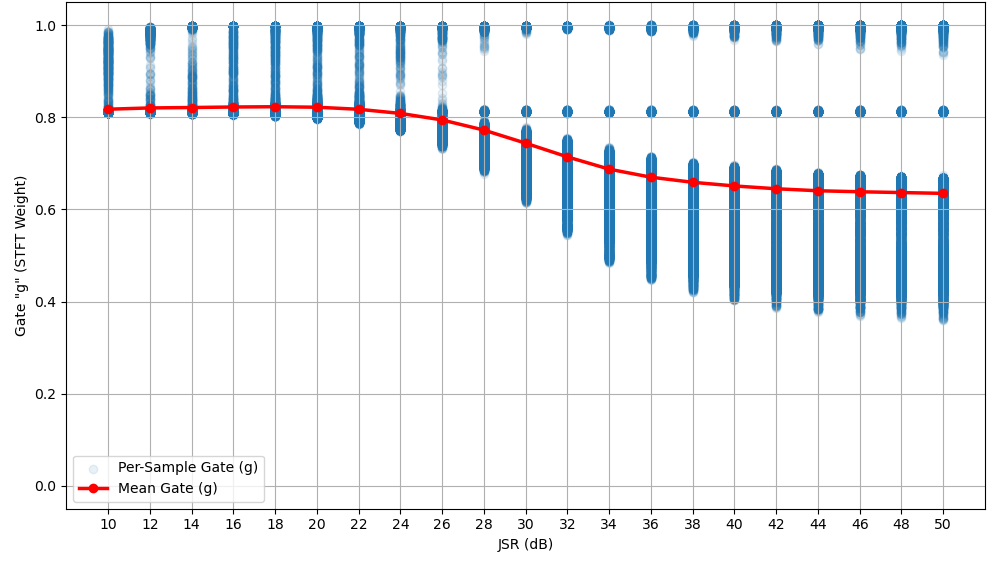}
    \caption{Internal dynamics of the primary gating mechanism ($g$) as a function of JSR.}
    \label{fig:gate_dynamics}
\end{figure}

Fig.~\ref{fig:gate_dynamics} visualizes the distribution of $g$ as a function of JSR. The observed trend aligns perfectly with the physical characteristics of the signal. At Low JSR (10--20 dB), the network assigns a high weight to the STFT branch (mean $g \approx 0.82$). This indicates that when the signal is submerged in noise, the model correctly identifies that visual spectral patterns are more reliable than phase-corrupted IQ data. As the signal quality improves in the High JSR (40--50 dB) range, the gate value decreases to a mean of approximately $0.63$. This shift signifies an increased reliance on the IQ branch (weighted by $1-g$), allowing the model to leverage precise phase constellations for fine-grained classification.

\begin{figure}[htbp]
    \centering
    \includegraphics[width=0.95\linewidth]{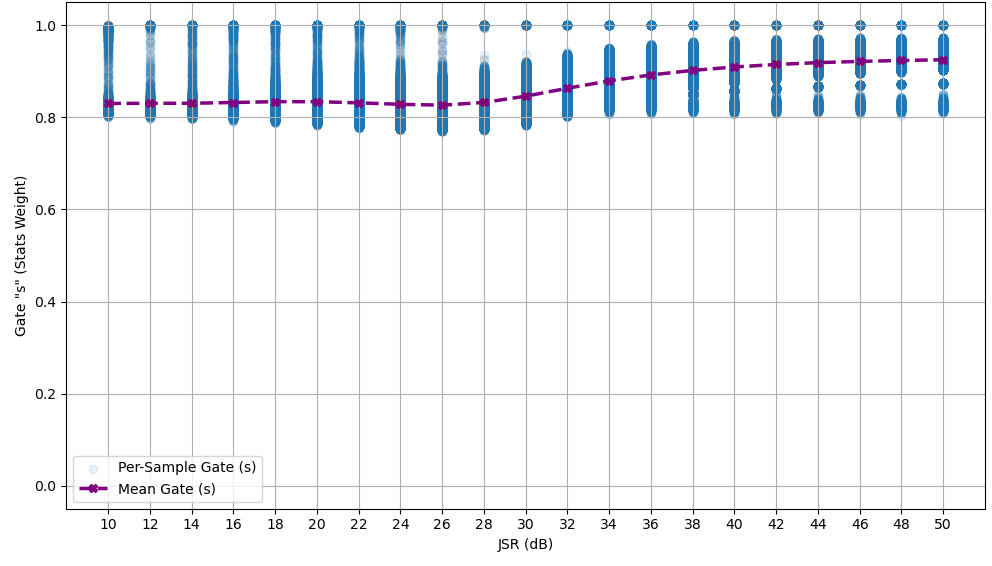}
    \caption{Dynamics of the auxiliary statistical gate $s$ versus JSR.}
    \label{fig:gate_s_dynamics}
\end{figure}

To complement the analysis of the primary modality switching, we further investigate the behavior of the auxiliary injection gate $s$, which controls the contribution of explicit statistical descriptors. Fig.~\ref{fig:gate_s_dynamics} illustrates the distribution of the gate weight $s$ across the JSR spectrum.

In contrast to the cross-over behavior observed in gate $g$, the auxiliary gate $s$ exhibits a consistently high activation level, with the mean weight remaining above 0.82 across the entire operational range. Unlike the primary gate which acts as a switch, this persistent activation confirms that statistical descriptors, such as Kurtosis and PAPR, provide a fundamental and robust baseline ``fingerprint'' for classification, maintaining reliability across all power levels. Furthermore, a slight upward trend is observed in the high-JSR regime (rising to approximately 0.92 at 50 dB). This phenomenon aligns with signal processing theory: as the interference power increases relative to the noise floor, the statistical estimators become more accurate and distinct, prompting the network to assign even greater confidence to this auxiliary branch to fine-tune the decision boundaries.

\subsection{Class-Specific Performance and Confusion Analysis}

To provide a deeper insight into the model's classification capabilities, we analyze the CM and the progression of F1-Scores.

\begin{figure}[htbp]
    \centering
    \includegraphics[width=0.75\linewidth]{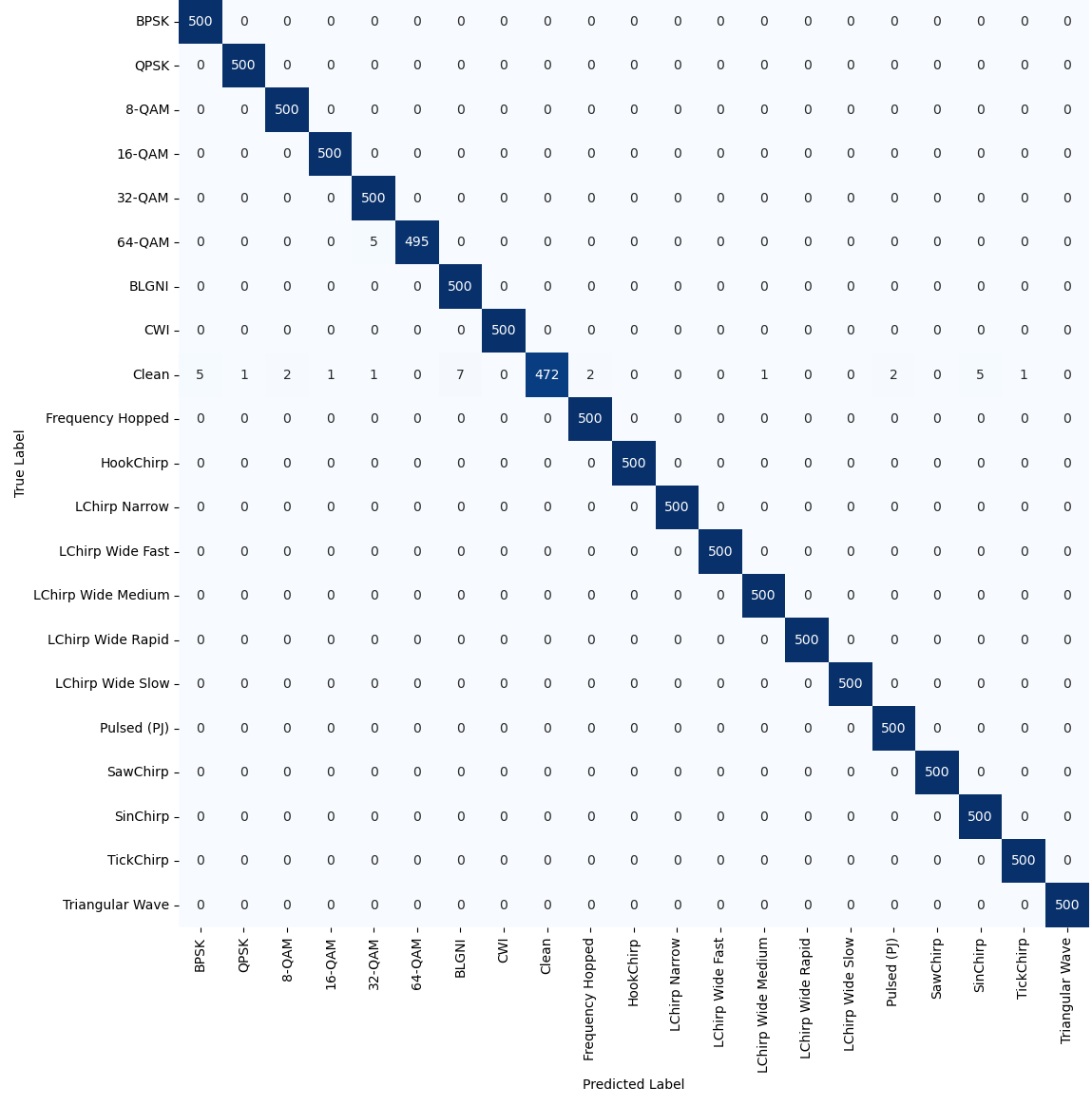}
    \caption{Confusion Matrix evaluated at JSR = 40 dB.}
    \label{fig:cm_40db}
\end{figure}

Fig.~\ref{fig:cm_40db} displays the CM at JSR = 40 dB, representing the model's performance in high-quality signal environments. The matrix exhibits a near-perfect diagonal structure, confirming the robustness of the proposed architecture. Specifically, JSR-GFNet successfully resolves the inter-class ambiguities among high-order digital modulations (16-QAM, 32-QAM, 64-QAM) that typically suffer from severe confusion at lower JSRs. The clear diagonal indicates that the model has effectively learned to distinguish all 21 interference types by leveraging the precise phase information from the IQ stream.

To further quantify the recovery from noise-induced confusion, Table~\ref{tab:f1_score} summarizes the F1-Score progression for representative challenging classes. It is evident that for high-order modulations like 64-QAM, the performance leaps from a poor 0.45 at 20 dB to a perfect 1.00 at 40 dB.

\begin{table}[htbp]
    \caption{F1-Score Progression for Challenging Jamming Classes}
    \label{tab:f1_score}
    \centering
    \begin{tabular}{@{}lccccc@{}}
    \toprule
    \textbf{Jamming Class} & \textbf{10 dB} & \textbf{20 dB} & \textbf{30 dB} & \textbf{40 dB} & \textbf{50 dB} \\ \midrule

    QPSK    & 0.35 & 0.58 & 1.00 & 1.00 & 1.00 \\
    16-QAM & 0.12 & 0.41 & 0.95 & 1.00 & 1.00 \\
    64-QAM & 0.10 & 0.45 & 0.98 & 1.00 & 1.00 \\ \midrule

    LChirp Wide & 0.88 & 1.00 & 1.00 & 1.00 & 1.00 \\
    SawChirp    & 0.45 & 1.00 & 1.00 & 1.00 & 1.00 \\
    CWI         & 0.85 & 1.00 & 1.00 & 1.00 & 1.00 \\ \bottomrule
    \end{tabular}
\end{table}

\subsection{Per-Class Sensitivity Analysis}

\begin{figure}[htbp]
    \centering
    \includegraphics[width=0.95\linewidth]{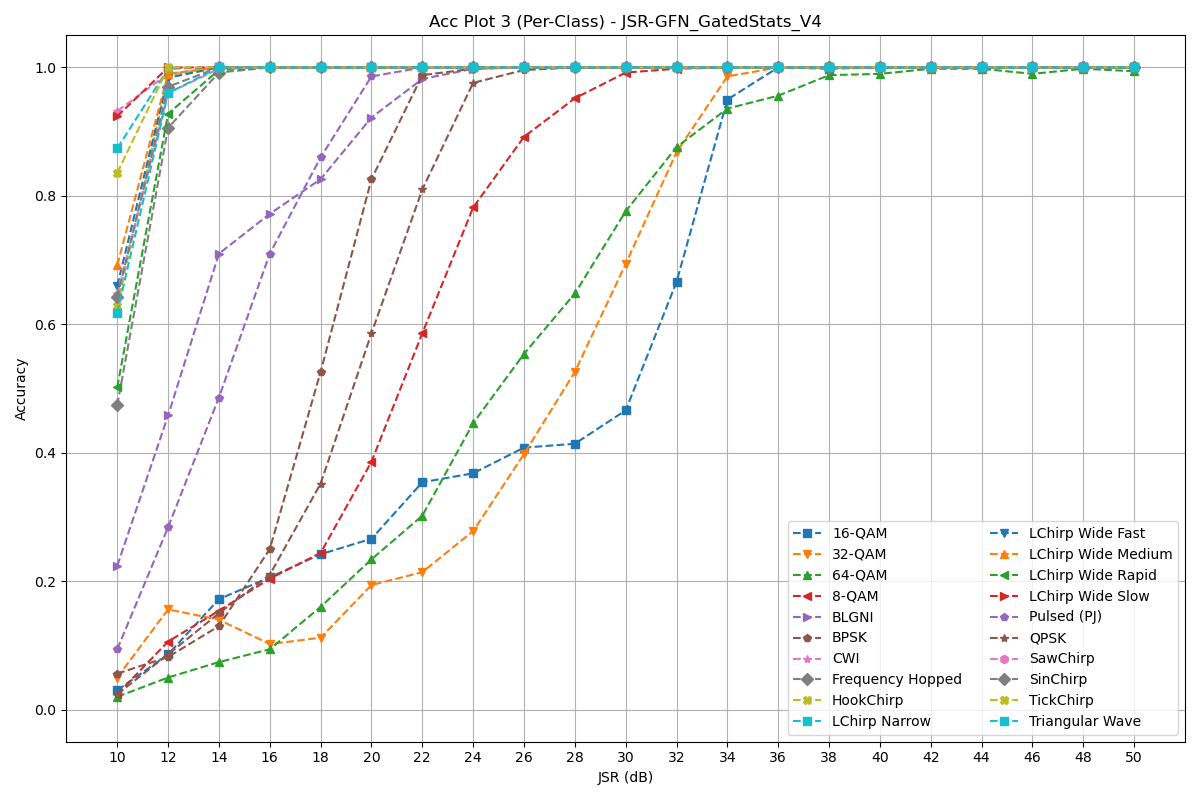}
    \caption{Per-class classification accuracy versus JSR, illustrating the distinction between Early Risers and Late Risers.}
    \label{fig:per_class_acc}
\end{figure}

Finally, Fig.~\ref{fig:per_class_acc} presents the accuracy curves for all 21 classes, allowing us to categorize the interference types based on their sensitivity to noise. We observe two distinct behaviors. First, the Early Risers, denoting robust classes such as CWI, Triangular Wave, and LChirp Wide, reach saturation ($>$95\% accuracy) at very low JSRs (12--14 dB). Their sparsity in the time-frequency domain makes them easily detectable by the STFT branch even in heavy noise. Second, the Late Risers, or sensitive classes including wideband signals like BLGNI and high-order QAMs, require significantly higher power (24--28 dB) to achieve reliable classification. The performance lag of these classes highlights the inherent physical difficulty of distinguishing them from the thermal noise floor.

\subsection{Computational Complexity and Efficiency Analysis}

To assess the feasibility of deploying JSR-GFNet on resource-constrained cognitive receivers, we benchmarked its computational complexity against baseline architectures. Evaluations were conducted on an NVIDIA 5060 Ti platform, measuring model size (Params), arithmetic intensity (FLOPs), and inference latency.

As detailed in Table~\ref{tab:complexity}, the proposed architecture demonstrates a superior efficiency-accuracy trade-off. By leveraging lightweight depth-wise separable convolutions in the spectral branch and efficient 1D operations in the temporal branch, JSR-GFNet requires only 6.85 million parameters---a reduction of approximately $15\times$ and $7\times$ compared to ViT-B-16 and AlexNet, respectively. In terms of computational cost, it achieves the lowest complexity with 0.48 GFLOPs. Despite the dual-stream fusion overhead, the inference latency remains at 3.12 ms, comparable to ResNet-18 and significantly faster than the Transformer-based approach. These metrics confirm that the proposed dynamic gating mechanism introduces negligible overhead, validating the model's suitability for near-real-time interference monitoring in next-generation avionics.

\begin{table}[htbp]
    \caption{Comparison of Computational Complexity and Inference Latency}
    \label{tab:complexity}
    \centering
    \begin{tabular}{@{}lccc@{}}
    \toprule
    \textbf{Model Architecture} & \textbf{Params (M)} & \textbf{FLOPs (G)} & \textbf{Latency (ms)} \\ \midrule
    AlexNet & 51.10 & 0.71 & \textbf{1.24} \\
    ResNet-18 & 11.69 & 1.82 & 2.45 \\
    ViT-B-16 & 106.36 & 17.58 & 8.36 \\
    \textbf{JSR-GFNet (Ours)} & \textbf{6.85} & \textbf{0.48} & 3.12 \\ \bottomrule
    \end{tabular}
\end{table}

\section{Conclusion}
\label{sec:conclusion}

This paper has presented JSR-GFNet, a physics-aware deep learning framework designed to address the fundamental limitations of existing GNSS interference classifiers in dynamic electromagnetic environments. By identifying the complementary nature of phase-sensitive time-domain features and texture-rich time-frequency representations, we constructed a multi-modal architecture that transcends the performance boundaries of single-stream baselines. The core contribution lies in the proposed decoupled dynamic gating mechanism, which functions as an autonomous reliability arbiter. This mechanism allows the network to adaptively shift its attentional focus from spectral energy integration in noise-limited low-JSR regimes to fine-grained phase constellation analysis in high-SNR scenarios.

Extensive evaluations on the comprehensive CGI-21 dataset demonstrate that JSR-GFNet not only achieves state-of-the-art classification accuracy across a wide dynamic range (10--50 dB) but also successfully resolves the critical ``feature degeneracy'' problem associated with high-order digital modulations—a task where traditional spectrogram-based CNNs fail. Furthermore, the interpretability analysis reveals that the learned gating behaviors align consistently with signal processing intuition, providing a degree of transparency essential for safety-critical aerospace applications. With its low computational latency and robust generalization capabilities, JSR-GFNet establishes a resilient foundation for the deployment of cognitive anti-jamming strategies in next-generation GNSS receivers.

\appendices

\section{Fundamental Ambiguity of Magnitude-Only Representations}
\label{app:ambiguity}

Let $x(t) \in L^2(\mathbb{R})$ denote a complex baseband signal. Its STFT with respect to a window function $g(t)$ is formally defined as:
\begin{equation}
    V_x(t,f) = \int_{-\infty}^{\infty} x(\tau) g(\tau-t) e^{-j2\pi f \tau} d\tau.
    \label{eq:stft_def}
\end{equation}
The standard spectrogram representation corresponds to the mapping $\mathcal{S}$:
\begin{equation}
    \mathcal{S}: x(t) \mapsto |V_x(t,f)|^2.
    \label{eq:spectrogram_map}
\end{equation}

We assert that $\mathcal{S}$ is fundamentally non-injective. Consider the set of all signals that map to the same spectrogram $P(t,f) \in \mathbb{R}_{\ge 0}$. Any signal $x(t)$ in the pre-image $\mathcal{S}^{-1}(P)$ satisfies $|V_x(t,f)| = \sqrt{P(t,f)}$, but its phase $\Phi_x(t,f) = \arg(V_x(t,f))$ remains unconstrained, subject only to the reproducing kernel Hilbert space constraints of the STFT range. Specifically, for any signal $x_1(t)$, there exists a class of distinct signals $\{x_k(t)\}$ such that:
\begin{equation}
    |V_{x_k}(t,f)|^2 \equiv |V_{x_1}(t,f)|^2,
    \label{eq:equivalence_condition}
\end{equation}
where the signals are related by phase transformations that preserve the magnitude envelope.

In the context of interference classification, this implies that two distinct stochastic processes---e.g., a high-order QAM signal $x_{\text{QAM}}(t)$ and a shaped Gaussian noise process $x_{\text{noise}}(t)$---can exhibit identical expected power spectral densities:
\begin{equation}
    \mathbb{E}\left[|V_{x_{\text{QAM}}}(t,f)|^2\right] = \mathbb{E}\left[|V_{x_{\text{noise}}}(t,f)|^2\right].
    \label{eq:stat_ambiguity}
\end{equation}
Consequently, a classifier $\mathcal{C}$ operating solely on the domain of $\mathcal{S}$ is theoretically bounded by an irreducible error when distinguishing between classes that form such an equivalence class. This formalizes the necessity of retaining phase information to resolve the ambiguity inherent in magnitude-only time-frequency distributions.

\section{Interference Power–Dependent Reliability of Signal Representations}

We provide a theoretical justification for adaptive, JSR-dependent fusion by modeling the IQ-domain and STFT-domain representations as two heterogeneous observation channels with distinct noise sensitivities. Let $z_I=g_I(x)+n_I$ and $z_S=g_S(x)+n_S$ denote features extracted from the IQ stream and the spectrogram stream, respectively, where the effective noise terms $n_I$ and $n_S$ depend on the JSR. Due to phase noise amplification and thermal noise dominance, the variance of phase-sensitive IQ features increases rapidly as JSR decreases, whereas the STFT representation benefits from time--frequency energy integration, yielding comparatively stable statistics in low-JSR regimes. Conversely, at high JSRs, the IQ stream preserves fine-grained phase and constellation structure that is irreversibly lost in magnitude-only spectrograms. Let $\mathcal{R}_m(\mathrm{JSR})$ denote a measure of discriminative information for modality $m\in\{I,S\}$, such as the Kullback--Leibler divergence between class-conditional feature distributions. Under mild regularity assumptions, $\mathcal{R}_S(\mathrm{JSR})$ dominates $\mathcal{R}_I(\mathrm{JSR})$ at low JSR, while the reverse holds at sufficiently high JSR for phase-modulated interference. Consider a linear fusion rule $z=\alpha z_I+(1-\alpha)z_S$. Minimization of the classification risk yields an optimal fusion weight $\alpha^\star$ that depends on the relative discriminative power of the two modalities and typically satisfies
\begin{equation}
\alpha^\star(\mathrm{JSR})\propto
\frac{\mathcal{R}_I(\mathrm{JSR})}
{\mathcal{R}_I(\mathrm{JSR})+\mathcal{R}_S(\mathrm{JSR})}.
\end{equation}
Since $\mathcal{R}_I(\mathrm{JSR})$ and $\mathcal{R}_S(\mathrm{JSR})$ exhibit opposite monotonic trends with respect to JSR, no constant fusion weight can be optimal across all operating conditions. This demonstrates that static fusion strategies are inherently suboptimal in environments with variable interference power and establishes the theoretical necessity of adaptive fusion mechanisms that respond to changes in JSR or its statistical proxies.

\end{document}